# The Benefits of Low Operating Voltage Devices to the Energy Efficiency of Parallel Systems

**Samuel Xavier-de-Souza[1], Eduardo A. Neves[1], Alex F. A. Furtunato[2], Luiz F. Q. Silveira[1], Kyriakos Georgiou[3], Kerstin I. Eder[3]**

[1]DCA-CT, Universidade Federal do Rio Grande do Norte, Natal-RN, Brazil, samuel@dca.ufrn.br
[2]Instituto Federal do Rio Grande do Norte, Natal-RN, Brazil,
[3]Department of Computer Science, University of Bristol, Bristol, United Kingdom

Programmable circuits such as general-purpose processors or FPGAs have their end-user energy efficiency strongly dependent on the program that they execute. Ultimately, it is the programmer's ability to code and, in the case of general purpose processors, the compiler's ability to translate source code into a sequence of native instructions that make the circuit deliver the expected performance to the end user. This way, the benefits of energy-efficient circuits build upon energy-efficient devices could be obfuscated by poorly written software. Clearly, having well-written software running on conventional circuits is no better in terms of energy efficiency than having poorly written software running on energy-efficient circuits. Therefore, to get the most out of the energy-saving capabilities of programmable circuits that support low voltage operating modes, it is necessary to address software issues that might work against the benefits of operating in such modes.

Multiple software layers are used to abstract away hardware details. Such abstraction enables the solution of complex problems while at the same time the programmer's productivity increases significantly. This comes with a caveat. Software developers often make poor programming choices due to the lack of understanding of how their code affects the usage of the various resources available on the hardware. Poor programming choices leading to a decrease in performance were not an issue since rapid hardware performance improvements were able to compensate for them. More recently, faster switching circuits are much harder to achieve on modern processors with high silicon density due to overheating and quantum effects. Moreover, with the IoT revolution, many systems are depended on limited or unreliable energy sources. These now require the software developer to put a significant effort in writing code that will meet both the energy and performance requirements for a large number of applications. This is a hard challenge as software developers lack the tools that will enable a resource aware software development [1].

To tackle the stagnation of single CPU performance, the microprocessor industry introduced multiprocessing on a single chip. The rationale is simple: if we cannot make them faster, let us make more of them and split the computation among them. With that, gradually, conventional sequential software, including the many layers that compose them, is giving place to parallel software. However, embracing this solution requires accepting a new challenge: write efficient parallel software. This is significantly harder than creating efficient sequential code. To make things worse, commonly, programmers are not equipped with the skills needed to create parallel software.

Addressing the software issues that prevent the benefits of low operating voltage modes reaching the end user might seem then to add up to the challenge of writing good parallel software. Nonetheless, parallel software could as well be an opportunity to employ low operating voltage devices more widely and efficiently. The reason for that stems from the following two statements. First, the number of existing layers in parallel software is much smaller than that of sequential software because of the discrepancy in the number of existing tools to write the two types of software. With fewer layers, the path to deliver device-level energy efficiency to the end user becomes much shorter. Second, parallel software allows another way to improve system performance other than increasing operating voltage and frequency: by increasing the number of processing elements, or cores in a multiprocessing chip, that carry out the computation. This is interesting because it creates an opportunity to operate computational systems at very low voltages while guaranteeing acceptable levels of performance. This work focus on demonstrating the latter statement.

If on one hand, parallel processing is a way to harvest the benefits of low operating voltage, on the other hand, low voltage is essential to allow parallel applications to run more energy efficiently. Being able to operate on lower voltages is necessary to accommodate operating frequency reductions that result from an increase in the number of cores. This interplay between number of cores and clock frequency is needed to sustain a given performance, which can be increased by more cores or higher voltages and decreased oppositely [2].

This work presents results of the analysis of mathematical models for performance and power consumption of multiprocessing chips that emphasize the need for low operating voltages. The models and the results are validated using simulation. They show that the better written the parallel software the larger the benefits of this low operating voltages.

The methodology presented here to improve software energy efficiency takes advantage of the quadratic relationship between power and voltage,

$$P = k_1 \cdot a \cdot C \cdot V^2 \cdot F + I_{leak} \cdot V, \qquad (1)$$

where $C$ is the total capacitance, $V$ is the supply voltage, $F$ is the operating frequency, $I_{leak}$ is the leakage current, $k_1$ is a proportionality constant, and $P$ is the power of a CMOS circuit. For a large region of the devices' operating point, this relationship is close to cubic due to the quasi-linear relationship between the operating voltage and the maximum allowed frequency in this region,

$$F_{max} = k_2 \cdot (V - V_{th})^h / V \qquad (2)$$

where $V_{th}$ is the threshold voltage, and $h$ is a technology dependent constant commonly assumed to be $h = 1.5$. Classical Voltage and Frequency Scaling (VFS) and Dynamic VFS (DVFS) employ the same principle to save energy. Additional to that, the approach presented here also considers the scalability of the computation w.r.t. the number of cores employed in a workload in order to scale down voltage and frequency.

The operating frequency must be chosen such that it allows sustaining performance with an increase in the number of cores. For that, the performance increase needs to be compensated by an equivalent reduction of clock frequency. This performance increase is given by the speedup $S_p$ for $p$ processing cores running at a given frequency $F$,

$$S_p(F) = T_s(F)/T_p(F) \qquad (3),$$

where $T_s(F)$ and $T_p(F)$ are the sequential and parallel execution times, respectfully.

The reference performance is assumed to be that of a single core running at maximum voltage, $V_s$, and frequency, $F_s$, which results in an execution time $T_s$. Hence, the frequency $F_p$ to be employed to the p cores for a sustained single core performance at clock frequency $F_s$ is given by

$$F_p = S_p \cdot F_s / T_r \qquad (4),$$

where $T_r$ is the ratio between the target parallel execution time $T_p$ and the single-core reference execution time $T_s$, such that a $T_r = 0.5$ represents a twice higher performance target compared to the single-core reference. The minimum voltage $V_p$ needed to allow the system to operate with a clock frequency $F_p$ is given by solving (2) for $V$ with $F_{max} = F_p$. The energy consumption of a program can then be obtained by combining (1), (2), (3), and (4), as follows:

$$E = p \cdot P \cdot T_p = p \cdot (k_1 \cdot a \cdot C \cdot V_p^2 \cdot F_p + I_{leak} \cdot V \cdot p) \cdot T_s(F_s) \cdot S_p \cdot F_s / F_p. \qquad (5)$$

Equation (5) assumes that a given multiprocessing chip with a number of cores larger than $p$ is employed. The power dissipated by the cores not used is assumed to be zero or neglectable by the use of power gating. Equation (5) allows the following methodology to obtain the energy consumption: *i*) given $k_1$, $k_2$, $C$, $V_{th}$, $h$, and $T_r$, assume $a$ and $I_{leak}$ to be constant, and $S_p$ to be known for any $p$ of interest; *ii*) obtain $F_p$ from (4) for any $p$ of interest; *iii*) calculated $V_p$ from (2) for the calculated $F_p$; *iv*) calculate the power dissipated by each core using (1); *v*) use (5) to calculate the energy consumption.

The modeling of (5) was analyzed numerically and using two tools: the Sniper multiprocessor simulator [3], developed by Intel®; and the Mcpat power modeling [4], developed by HP®. The architecture of the simulated multiprocessor chip was the same as in [5], containing 16 tiles of 4 cores, totaling 64 processing cores. The application simulated was BlackScholes, extracted from the Parsec parallel benchmark [6]. The parameters used in the numerical analysis and in the simulation are presented in Table 1.

The main analytical results are presented in Fig. 1. The energy consumption was plotted for different performance targets, modeled by $T_r$, and different types of parallel applications, modeled by their parallel fraction *f*. Fig. 1 also shows the values of the supply voltage for different types of parallel applications with a performance target $T_r = 0.25$, i.e. a performance four times higher than the single-core performance at maximum frequency. These plots show that an increase in the number of cores used to carry out the computation not only could reduce the energy consumption by orders of magnitude but also requires a significant reduction in the operating voltage. Also, the reductions in energy and voltage are larger for applications that are more parallel, emphasizing the need for well-written parallel codes. Parallel codes with smaller *f* have the optimal point of energy consumption at higher values of energy and voltage and lower numbers of processing cores.

The simulation results that validate the trends of the analytical modeling are depicted in the left most plot of Fig. 1. The application BlackScholes ran with $T_r = 1$ and the energy measurements follow the respective analytical curve.

In summary, this work presents results from analytical models validated by state-of-the-art simulation showing that well-written parallel software allows large reductions in energy consumption by the use of low operating voltage. It has been shown that codes with larger parallel fractions can take more advantage of lower voltages to enable larger reductions. Conclusively, employing parallel software in systems built with low operating voltage devices enables large energy savings, which are lager for lower operating voltages.

Acknowledgment:
This work was supported in part by the Royal Society-Newton Advanced Fellowship granted to S. Xavier-de-Souza (Award #13303 NA160108) and by the CNPq, Brazil.

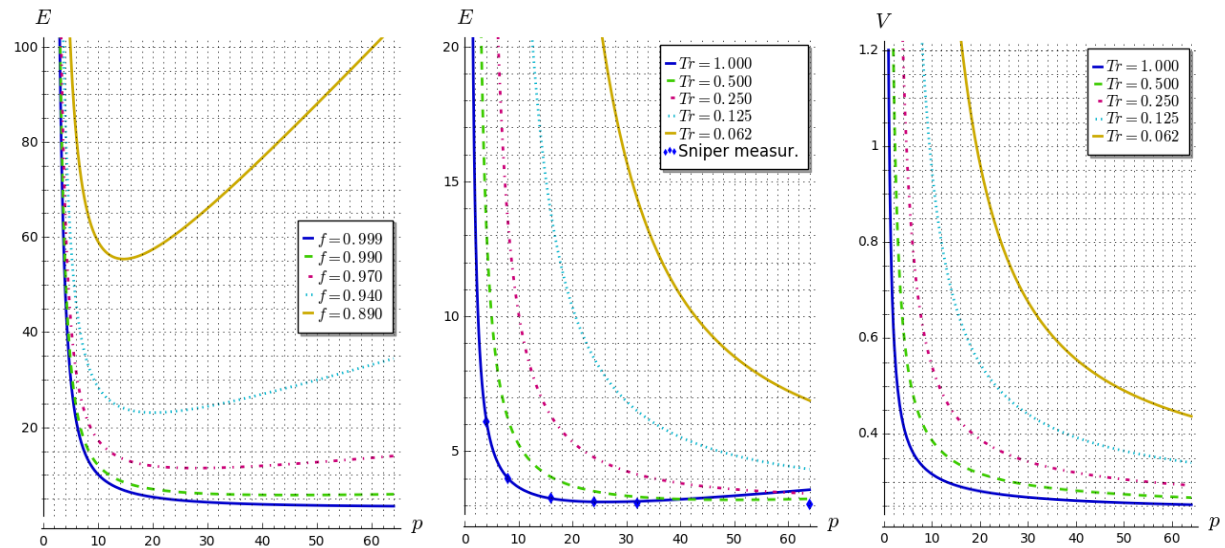

Figure 1. Energy consumption $E$ and supply voltage $V$ for different numbers of processing cores $p$.

| Parameter | Symbol | Value |
|---|---|---|
| Constant including activity factor and total capacitance of a processing core | $k_1 \cdot a \cdot C$ | $1.06 \times 10^{-8}$ |
| Constant representing the total leakage current of a processing core | $I_{leak}$ | $7.97 \times 10^{-2}$ A |
| Proportionality constant | $k_2$ | $4.02 \times 10^{-9}$ |
| Threshold voltage | $V_{th}$ | 0.23V |
| Technology-dependent minimum delay constant | $h$ | 1.5 |
| Reference performance single-core frequency | $F_s$ | 3.2 GHz |
| Reference performance single-core voltage | $V_s$ | 1.2V |

Table1. Parameters used for the analytical and simulation results.